\documentstyle[12pt]{article}

\renewcommand{\Large}{\large}
\catcode `@=11 \@addtoreset{equation}{section} \catcode `@=12

\begin{document}

\def\del{\partial}
\def\wf{\mbox{$W_\infty^f$}}
\def\be{\begin{equation}}
\def\ee{\end{equation}}
\def\ba{\begin{array}{l}}
\def\ea{\end{array}}
\def\A{\mbox{$\bar A$}}
\def\M{\mbox{$\overline M$}}
\def\eq#1{(\ref{#1})}
\def\D{\mbox{$\cal D$}}
\def\psid{\mbox{$\psi^\dagger$}}
\def\tr{{\rm tr}}
\def\Tr{{\rm Tr}}
\def\aohat{\mbox{$\widehat{ A_0}$}}
\def\l{\lambda}
\def\rhot{\mbox{${\widetilde \rho}$}}
\def\U{\mbox{$\cal U$}}
\def\fr{{\rm frac}}
\def\Hc{\mbox{${\cal H}_c$}}
\def\Hct{\mbox{${\tilde {\cal H}_c}$}}
\def\Pit{\mbox{${\widetilde \Pi}$}}
\def\Vac{\mbox{$| \vec{ n_F } \rangle $}}
\def\xib{\mbox{$ \bar \xi $}}

\renewcommand\arraystretch{1.5}

\begin{flushright}
TIFR-TH-94/36
\end{flushright}
\begin{center}
\vspace{3 ex}
{\Large\bf
LEADING LARGE $N$ MODIFICATION OF QCD$_2$
}\\
\vspace{1 ex}
{\Large\bf
ON A CYLINDER
}\\
\vspace{1 ex}
{\Large\bf
BY DYNAMICAL FERMIONS
}\\
\vspace{10 ex}
Avinash Dhar$^*$, Gautam Mandal and Spenta R. Wadia\\
Tata Institute of Fundamental Research \\
Homi Bhabha Road, Bombay 400 005, INDIA \\
\vspace{20 ex}
\bf ABSTRACT\\
\end{center}
\vspace{2 ex}
We consider 2-dimensional QCD on a cylinder, where space is a circle.
We find the ground state of the system in case of massless quarks in a
$1/N$ expansion. We find that coupling to fermions nontrivially
modifies the large $N$ saddle point of the gauge theory due to the
phenomenon of `decompactification' of eigenvalues of the gauge
field. We calculate the vacuum energy and the vacuum expectation value
of the Wilson loop operator both of which show a nontrivial dependence
on the number of quarks flavours at the leading order in $1/N$.
\thispagestyle{empty}
\vfill
\hrule
\vspace{1 ex}
\noindent{\small e-mail: adhar, mandal, wadia@theory.tifr.res.in}
\hfill\break \noindent{\small
$^*$Address after October 1, 1994: Theory Division, CERN, CH-1211,
Geneve 23, Switzerland.}
\eject

\setcounter{section}{-1}
\section{Introduction and Summary}

The large $N$ expansion provides a valuable tool for obtaining
qualitative insight into gauge theories \cite{BOOK}.  The $1/N$
expansion in gauge theories is based on the discovery by 'tHooft
\cite{THOOFT}\ that for $SU(N)$ gauge theories with fundamental
fermions the $N$ dependence of a typical vacuum diagram $E_{H,L}$ for
large $N$ is
\be
E_{H,L} \propto  N^2 ({ 1 \over N^2 })^H ( { 1 \over N } )^L
\label{thooft}
\ee
where $H$ is the number of handles and $L$ the number of holes
(fermion loops) in the diagram. It is easy to see that the dominant
contributions to the vacuum energy occurs when $H=L =0$ and that it is
proportional to $N^2$. This simply reflects the fact that there are
$N^2$ gluons in the theory. The fermionic contribution to the vacuum
energy, on the other hand, must have at least one fermion loop, and
the leading contribution is $E_{0,1} \propto N$.  This, again,
reflects the fact that there are $N$ fermions in a fundamental
representation.  The $N^2$-dependence of the vacuum energy has been
explicitly calculated in soluble large $N$ matrix models
\cite{MATRIX}\ which, like gauge theories, also possess $N^2$ degrees
of freedom.  On the other hand, two-dimensional QCD (QCD$_2$) on the
plane, which is another soluble model at large $N$ \cite{THOOFT2},
possesses no dynamical gauge degrees of freedom and has vacuum energy
$\propto N$ coming from the $N$ fermions. See also the more recent
works on pure QCD$_2$ \cite{QCD2YM}\ and on QCD$_2$ with fermions
\cite{QCD2,DMWQCDA}.

In the present paper we continue \cite{DMWQCDA}\ our study of large
$N$ QCD$_2$ on a cylinder. Here space is a circle and hence the gauge
field does not decouple. After fixing gauge appropriately the gauge
field can be described by the $N$ eigenvalues of $A_1$ which satisfy
fermi statistics. The standard wisdom, based on Eqn. \eq{thooft}\ and
the discussion in the last paragraph, would suggest the following
two-step procedure for solving the theory at large $N$ --- (1) to
solve the pure gluon theory first (the large $N$ saddle point of this
is easy to determine and is described by a constant density of
eigenvalues; see remarks before \eq{wilsonvalueym} below) and (2)
treat the fermion dynamics subsequently as fluctuations in the fixed
external gauge field background determined in step (1). In this paper
we explicitly show that such a procedure is incorrect. The reason for
this, basically, is that in presence of quarks there are
gauge-invariant operators which (unlike the Wilson loop operator) are
{\em not periodic} functions of the eigenvalues and this effectively
leads to a noncompact range of the eigenvalues unlike in the pure
Yang-Mills theory. This is the phenomenon of `decompactification'
described in great detail in \cite{DMWQCDA}.  It is clear that a
constant density of eigenvalues is not normalizable in a noncompact
domain. Indeed we find that the quarks lead to a harmonic oscillator
potential for the eigenvalues, resulting in expressions for physical
quantities that are nontrivially different from those of pure
Yang-Mills theory at the leading $N$ order. We present our main
results below.

The vacuum energy in our theory is given by (see Eqn. \eq{enf} below)
\be
\bar E_0 =  N^2 [ \sqrt{n_f} { \bar g \over 2 \pi } + o(1/N) ]
\ee
where $ n_f $ is the number of quark flavours and $ {\bar g} = g
\sqrt{ N}$ as usual denotes the scaled coupling constant. The vacuum
expectation value of the Wilson loop operator $W_m =(1/N) \Tr U^m, \;
U = \exp [i \int_0^L A_1 dx]$ is given by (see Eqn. \eq{wilsonvalue})
\be
\langle W_m \rangle =  2 (-1)^m  (1 + { \del^2 \over \del {x_m}^2 } )
J_0(x_m) + o(1/N), \quad x_m \equiv 2 \pi^{3/4}
\alpha { m \over n_f^{1/4}}, \; \alpha = \sqrt{ \bar g \bar L / 2 \pi}
\ee
In the above $\bar E_0 \equiv E_0 \sqrt{N}$ and $\bar L \equiv L
\sqrt{N}$ denote an additional $N$-scaling necessary in order to have
a well-defined large $N$ limit in our theory. The expression for the
eigenvalue density is given by Eqn. \eq{rhonf}.

Note the nontrivial dependence of the above expressions on the number
of quark flavours which clearly shows that adding quarks changes the
leading large $N$ result. For comparison, note that the vacuum
expectation value of the Wilson loop operator in pure Yang-Mills theory
in two dimensions is (see \eq{wilsonvalueym}\ below)
$ \langle W_m \rangle_{YM} = \delta_{m,0} + o(1/N).$

This paper is organized as follows. In Sec. 1 we write down the
largrangian, the path integral and the hamiltonian for QCD on a
two-dimensional cylinder. We follow the notation and results of
\cite{DMWQCDA}. In Sec. 2 we find the ground state of the theory
in the $1/N$ expansion and reduce the calculation of the vacuum energy
to a problem of $N$ interacting fermions in a harmonic oscillator
external potential. The solution of the latter problem is presented in
Sec. 3 in the $1/N$ expansion which allows us to calculate the vacuum
energy of the full system. We find in the process that we need to
scale both the radius of the cylinder and the time in a certain way
(see Eqns. \eq{lengthscaling}\ and \eq{tscaling}) to have a
well-defined large $N$ limit.  Besides the vacuum energy we also
calculate the vacuum expectation value of the Wilson loop operator.
In Sec. 4 we conclude with some comments about possible relevance to
four dimensions.

\section{The Action and the Hamiltonian}
We consider the gauge group $U(N)$. As usual we denote the gauge
fields, which are hermitian matrices, by $A^{ab}_\mu$ where $\mu =
0,1$ is the Lorentz index and $a,b = 1,2, \ldots, N$ are colour
indices. The fermions are denoted by $\psi^a_{i\alpha}$ where $i=1,2,
\ldots,n_f$ is the flavour index and $\alpha= {\pm 1} $
is the dirac index.  We consider one space and one time dimension
where the space dimension is a circle of length $L$.

The theory is described by the following path-integral
\be
\ba
Z = \int \D A_0(x,t) \D A_1(x,t) \D\psi(x,t) \D\psid(x,t)
\exp[iS(A_0,A_1, \psi,\psid)] \\
S = \int_0^T dt \int_0^L dx \big(- {1\over 4} \tr F_{\mu\nu}
F^{\mu\nu} + \bar\psi \gamma^\mu (iD_\mu) \psi - m\bar\psi
\psi ) \\
F_{01} \equiv E = \del_t A_1 - \del_x A_0 + ig [A_0, A_1],\quad
D_\mu = \del_\mu + ig A_\mu
\ea
\label{pathintegral}
\ee
In this paper we will only consider the case of massless quarks,
$m=0$.

Let us fix the gauge
\be
A_1(x,t) =  \hbox{Diag}[\l_a(t)]
\label{gaugechoice}
\ee
The gauge-fixed path-integral becomes \cite{DMWQCDA}
\be
Z = \int \D \l(t) \Delta_P(\l(0)) \Delta_P (\l(T)) \D \psi(x,t) \D \psid(x,t)
\exp[i (S_{YM} + S_F)]
\label{gaugefixedz}
\ee
where
\be
\ba
S_0 = \int_0^T dt\, {L\over 2} \sum_a (\del_t \l_a)^2 \\
S_F = \int_0^T dt\int_0^L dx \big[\psid i\del_t \psi + \psid
\gamma^5 (i\del_x -g \l) \psi \big] + S_{coul} \\
S_{coul} = - \sum_{a,b} \int_0^T dt \int_0^L dx\int_0^L dy \rho_{ab}(x)
\rho_{ba}(y) K_{ab}(x-y)
\ea
\label{action}
\ee
The kernel $K_{ab} (x)$ is given by (for $x \in [-L, L]$)
\be
\ba
K_{ab}(x) =  (g^2/4) e^{igx(\l_a - \l_b)} \big[ L /( 2 \sin^2
{ \pi (\l_a - \l_b) \over \l_0} ) - ix {\rm cot}
{ \pi (\l_a -\l_b ) \over \l_0 }
- |x| \big], \; a \neq b
\\
K_{aa}(x) =  (g^2/4) \big[ L / 6 - |x| + x^2/ L \big]
\ea
\label{kernelab}
\ee
The action in \eq{action}\ is invariant under the ``large'' gauge
transformations \cite{DMWQCDA,MANTON,LARGEGAUGE}
\be
\ba
\psi_a(x,t) \to \Omega_{ab}(\vec m, x)\psi_b(x,t), \quad
\Omega_{ab}(\vec m, x) \equiv \exp[-igxm_a \l_a]
\\
\l_a(t) \to \l^\Omega_a(t) = \l_a (t) + m_a \lambda_0
\ea
\label{large}
\ee
where
\be
\lambda_0 \equiv \frac{2\pi}{gL}
\label{lambdaperiod}
\ee
The parameter $\l_0$ defines the range $[0, \l_0]$ of integration of
the eigenvalues $\l_a$ in \eq{gaugefixedz}. A simple way of
understanding the period \eq{lambdaperiod}\ is to consider the Wilson
loop operators
\be
W_m \equiv { 1 \over N }\Tr U^m, \; U \equiv \exp (i g\int_0^L A_1 dx)
\label{wilsondef}
\ee
which in the gauge \eq{gaugechoice}\ evaluate to
\be
W_m = {1 \over N} \sum_a \exp(i g L \l_a) \equiv { 1 \over N } \sum_a
\exp (i 2\pi \l_a/\l_0)
\label{gaugefixedwilson}
\ee
Mesonic operators \cite{DMWQCDA}\
or $H_F$ in \eq{hf}\ below, which involve gluons as well as quarks,
are not periodic under $\l_a =0 \to \l_a = \l_0$ {\em per se} but one
needs to simultaneously transform the quarks also according to
\eq{large}.  This ultimately leads to a decompactification of
eigenvalues in the effective theory of the gluons.

The factors $\Delta_P(\l(0))$ and $\Delta_P(\l(T))$ in the measure are
defined by
\be
\Delta_P(\l) \equiv  \prod_{a < b} \sin[\pi(\l_a - \l_b)/\l_0]
\label{sineq}
\ee
These factors in the measure imply that the initial and final
wavefunctions are completely antisymmetrized with respect to the
$\l_a$'s (since to start with they must be symmetric with repsect to
permutation of the $\l_a$'s on account of Weyl-symmetry).  This gives
rise to the well-known fermionic nature of these eigenvalues.

The hamiltonian corresponding to the action \eq{action}\ is given by
\be
H = H_{YM} +  H_F,
\label{hamiltonian}
\ee
where
\be
H_{YM} = \frac{1}{2L} \sum_a p^2_a,
\label{hym}
\ee
and
\be
\ba
H_F = H_{F,0} + H_{coul}
\\
H_{F,0}= \sum_a \int_0^L dx\,  \psi^{\dagger a} \gamma^5(-i\del_x
+ g \l_a)\psi^a
\\
H_{coul} = -\sum_{ab} \int_0^L dx \int_0^L dy\, \rho_{ab}(x) \rho_{ba}
(y) K_{ab}(x-y)
\ea
\label{hf}
\ee
The kernels $K_{ab}(x-y)$ are given by \eq{kernelab}.

The physical Hilbert space of the above theory satisfies the zero
charge condition (for each colour)  \cite{DMWQCDA}
\be
Q_a \equiv \int dx \rho_{aa}(x) = 0
\label{qeq}
\ee

\section{Ground State}
In this section we construct the ground state of the hamiltonian
\eq{hamiltonian}\ in the $1/N$ expansion. We do it  in two steps.
(a) We first consider the gauge field as external and discuss the
dynamics of the fermions for fixed $\l_a$'s. We show that for any
fixed $\l_a$'s the dirac sea built out of free fermions is the ground
state of this problem modulo $1/N^2$ corrections. (b) Next we
use this result to construct the ground state of the
full problem where both fermions and the eigenvalues are dynamical.

We should remark that the expectation value of $H$ in the full ground
state of the theory was already presented in \cite{DMWQCDA}. In the
discussion below we will allow some small overlap with \cite{DMWQCDA}\
for the sake of completeness.

\subsection{Fermion dynamics in an external background of gauge fields}

We start by discussing eigenstates of $H_{F,0}$ (eqn. \eq{hf}) in the
presence of fixed background values of $\{ \l_a \}$. We will include
the effect of $H_{coul}$ later on. For simpilcity we will also work
with $n_f=1$ at first.

Since $H_{F,0}$ is quadratic, the ground state is simply given by
filling the fermi sea according to the single-particle spectrum
\be
E^a_{n\alpha} = \frac{2\pi}{L} {\rm sgn}(\alpha) (n +
\frac{\l_a}{\l_0})
\label{diracspectrum}
\ee
Eqn. \eq{diracspectrum}\ is obtained by noting that  the dirac equation
$ [i\del_t + \gamma^5 (i\del_x - g\l_a) ] \psi^a_\alpha = 0 $
is solved by
\be
\psi^a_\alpha (x,t) = \sum_n \exp[-iE^a_{\alpha,n}t +
i\frac{2\pi n}{L} x] \psi^a_{\alpha,n}
\label{fermimode}
\ee
with $E^a_{\alpha,n}$ as in \eq{diracspectrum}. Here we have chosen
the convention $\gamma^0 = \sigma_1$ and $\gamma^1 = -i\sigma_2$, so
that $\gamma^5 \equiv \gamma^0 \gamma^1 = \sigma_3$.  $\alpha = \pm 1$
represent the right-, left-moving fermions respectively.

In filling the fermi sea, {\em a priori} the fermi levels could be
different for different chirality and different colour. However, as we
noted in \cite{DMWQCDA}, translation invariance of the vacuum demands
that the right fermi-momentum must be left fermi-momentum minus one
for each colour. Let us denote the right fermi-momentum for colour $a$
as $p^a_F = 2\pi n^a_F / L$. The fermi (dirac) sea is therefore the
state \Vac\ defined by
\be
\ba
\psi^a_{R,n} \Vac = 0, n > n^a_F \\
\psi^{\dagger a}_{R,n} \Vac = 0, n \le n^a_F \\
\psi^a_{L,n} \Vac = 0, n \le n^a_F \\
\psi^{\dagger a}_{L,n} \Vac = 0, n > n^a_F
\ea
\label{fermisea}
\ee
The fermion modes $\psi^a_{\alpha,n}$ are defined by \eq{fermimode}\
and satisfy the anticommutation relation
\be
\{  \psi^a_{\alpha, n}, \psi^{\dagger b}_{\beta, m} \} = {1 \over L}
\delta_{\alpha\beta} \delta_{mn}
\label{etar}
\ee

The reader might wonder how the fermi levels can be different for
different colours and still produce a colour-singlet state. Indeed the
question is further complicated by the fact that the fermi levels
$n^a_F$, having come from momentum labels of fermions, are
non-gauge-invariant; under the large gauge transformation \eq{large}\
they transform as
\be
n^a_F \to (n^a_F)^\Omega =  n^a_F - m_a
\label{fermilevelshift}
\ee

We will defer the detailed discussion of gauge-invariance of the
vacuum till Sec. 2.2. For the moment, let us check that the state
\Vac\ satisfies the constraint \eq{qeq}. Note that $Q_a = \sum_{\alpha
= R, L} Q_{\alpha, a}$ where $ Q_{\alpha, a} = L
\sum_{m= -\infty}^\infty \psi^{\dagger a}_{\alpha,m} \psi^a_{\alpha,
-m}$. Evaluating this on the state \Vac\ we get
\be
Q^a_R =  \sum_{m = -\infty}^{n^a_F} 1, \qquad
Q^a_L =  \sum_{m = n^a_F +1}^\infty 1
\label{chargedef}
\ee
The sums are divergent. By using the gauge-invariant exponential
regulator \cite{DMWQCDA}\ $\exp[ -\epsilon | m + \l_a/\l_0 |]$ we get
\be
Q^a_R(\epsilon) = {1 \over \epsilon} + {1 \over 2} + n^a_F +
{\l_a \over \l_0} + o(\epsilon), \quad
Q^a_R(\epsilon) = {1 \over \epsilon} - {1 \over 2} - n^a_F -
{\l_a \over \l_0} + o(\epsilon)
\label{chargevalue}
\ee
This leads to
\be
Q_a(\epsilon) = Q^a_R(\epsilon) + Q^a_L(\epsilon) = {2 \over \epsilon}
\label{totalcharge}
\ee
This is a divergent $c$-number term which can be removed by normal
ordering. Thus $: Q_a : = Q_a(\epsilon) - (2/\epsilon) = 0$. Note that
we have been able to achieve this without assuming any special values
of $n^a_F$.

\vspace{2 ex}

\noindent\underline{Construction of ground state of $H_F$}

\vspace{2 ex}

It is easy to see that \Vac\ is an eigenstate of the dirac part
$H_{F,0}$ of the hamiltonian \eq{hf}
\be
H_{F,0} \Vac = E_{F,0} \Vac
\ee
\be
E_{F,0} = g\l_0 \sum_a \sum_{m= -\infty}^\infty (m + \l_a/\l_0)
\tr_{d}(\gamma^5 Q^a_m)
\label{efo}
\ee
where
\be
Q^a_{m, \alpha\beta} \equiv L \langle \vec{n_F} | \psi^{\dagger
a}_{\alpha,m} \psi^a_{\beta, m} \Vac
= \theta(n^a_F - m) {1 + \gamma^5 \over 2} + \theta(m - n^a_F - 1)
{1 - \gamma^5 \over  2}
\label{qan}
\ee

Regarding the action of $H_{coul}$ on \Vac, it is more convenient to
use the alternative form
\be
H_{coul} = \sum_{\alpha, \beta} H_{\alpha\beta}
\ee
where
\be
H_{\alpha\beta} = {L \over 2 \l_0^2} \sum_{n= -\infty}^\infty {
\rho_{\alpha, ab, n}
\rho_{\beta, ba, -n}  \over [n + (\l_a - \l_b)/\l_0 ]^2 }
\label{hab}
\ee
In the above
\be
\rho_{\alpha, ab, n} =  \sum_{m= -\infty}^\infty \psi^{\dagger b}_{
\alpha, m} \psi^a_{\alpha, m+n}
\ee
It is a lengthy but straightforward calculation to show that
\be
[H_{RR} + H_{LL}] \Vac =  E_{coul} \Vac
\ee
where
\be
E_{coul} = {g \over 4 \pi \l_0} \sum_{a,b} \sum_{m,m'= -\infty}^\infty
{ \tr_d (Q^b_m Q^a_{m'}) \over [ m - m' + (\l_a - \l_b)/\l_0 ]^2 }
\label{ecoul}
\ee
where the $Q^a_n$ have been introduced above in \eq{qan}.

The left-right mixing terms $H_{LR} = H_{RL}$ take \Vac\ to an
orthogonal state (which is a linear combination of states with two
holes and two particles). If we treat $H_{LR}$ as a perturbation
term, the first order perturbation correction is zero since
\be
\langle \vec{n_F} | H_{LR} \Vac = 0
\ee
Second order perturbation theory gives the contribution
\be
E_{ L R } = { g \over 8 \pi^2 \l_0^3 } \sum_{ a, b } \sum_{ m = -
\infty }^\infty { m \over [ m + n^a_F - n^b_F + (\l_a - \l_b)/\l_0 ]^4
}
\label{elr}
\ee
For finite $N$ there is no reason to regard this as a perturbation.
However in the large $N$ limit we scale the coupling constant $g$ as
$g = \bar g/\sqrt{N}$ so that $\l_0 = \bar \l_0 \sqrt{N}$. This makes
$E_{LR}$ down by $1/N$ compared to $E_{coul}$ because of the two extra
powers of $\l_0$ in the denominator. This is actually the story with
conventional scaling of coupling constant. As we shall see in Sec. 3,
the scaling in our theory also involves scaling of the length $L$ of
the circle. This in fact brings down $E_{LR}$ by an additional factor
of $1/N$.

Combining the above results, we find that the dirac sea is an
eigenstate of $H_F$ to the leading order in $1/N$, with the energy
given by
\be
H_F \Vac =  E_F \Vac + o(1/N), \quad E_F =  E_{F,0} + E_{coul}
\label{efdef}
\ee
The sums in \eq{efo}\ and \eq{ecoul}\ are divergent. Using the
exponential regulator once again as in \eq{chargevalue}\ we get
\be
E_{F,0}(\epsilon) = g\l_0 \sum_a [-\frac{2} {\epsilon} -\frac{1}{12} +
V_{\rm reg} (\xi_a) + o(\epsilon)]
\label{efoe}
\ee
and
\be
E_{coul}(\epsilon) = {g \over 4\pi \l_0} \sum_{a,b} \big[
\frac{\pi^2} {\epsilon \sin^2 \frac{\pi}{\l_0}
(\xi_a -\xi_b) } + 2 (\ln \epsilon -1) + K ({\xi_a - \xi_b \over \l_0}) +
o(\epsilon \ln \epsilon) \big]
\label{ecoule}
\ee
where
\be
V_{\rm reg}(\xi_a) =  ({\xi_a \over \l_0} + {1\over 2})^2
\label{potential}
\ee
\be
K_{\rm reg}({\xi_a - \xi_b \over \l_0}) =
\left[ 2C +  \psi(w_{ab})
+ \psi(-w_{ab}) + w_{ab} \{ \psi'(w_{ab}) - \psi'(- w_{ab}) \}
\right]
\label{kernel}
\ee
Here $w_{ab} \equiv (\xi_a - \xi_b)/\l_0$ and $C$ is Euler's constant.
Also, $\psi(x) = (d/dx) \ln \Gamma(x) $ and $\psi'(x) =
\del_x \psi(x)$, $\Gamma(x)$ being the standard gamma-function.
Note that $E_F$ only depends on the gauge-invariant combination
\be
\xi_a \equiv \l_a + n^a_F\l_0
\label{xiadef}
\ee
The appearance of this variable is responsible for decompactification
of the eigenvalue $\l_a$. We will discuss this in more detail shortly.

A remark is in order here justifying our definition of the regularized
quantities $V_{\rm reg}$ and $K_{\rm reg}$. The divergent piece in
$E_{F,0}(\epsilon)$ is a constant and, because of the constraint that
the total number of eigenvalues is $N$, does not affect the dynamics.
In the grand canonical ensemble the equivalent statement is that such
a divergence can be cancelled by a simple additive renormalization of
the chemical potential term. Similar remarks can be made about the
$\ln(\epsilon)$ piece in $E_{coul}(\epsilon)$ and the constant finite
pieces in $E_{F,0}(\epsilon)$ and $E_{coul}(\epsilon)$. The
justification for ignoring the $1/(\epsilon \sin^2)$ piece in
$E_{coul}(\epsilon)$ is more subtle. The main point is that this
$1/\epsilon$ comes multiplied with the quadratic pole $(\xi_a - \xi_b
)^{ -2 } $ which is the only singularity of \eq{ecoule}\ in the limit
$\xi_a \to \xi_b$. If we take the same limit in the unregulated
expression \eq{ecoul}\ we find a quadratic pole with residue $ \sum_n
(Q^a_n)^2 = \sum_n Q^a_n = Q_a $. Here we have used $(Q^a_n)^2 = Q^a_n
\, \forall a,n$. Thus the $1/\epsilon$ in \eq{ecoule}\ can be identified
with the total charge which must vanish by \eq{qeq}.

Combining all this we get the following regularized expression for the
eigenvalue $E_F$
\be
\ba
E_F = g \lambda_0 \sum_a ( \xi_a/\lambda_0 + 1/2)^2
+ {g \over 4\pi \l_0} \sum_{a,b}  K_{\rm reg}( { \xi_a - \xi_b \over
\l_0})
\ea
\label{efvalue}
\ee
where $K_{\rm reg}$ is defined in \eq{kernel}.

In the above we have proved that the filled fermi sea is an eigenstate
of $H_F$ at leading $1/N$ order. How does one argue that it is
actually the {\em ground} state of the system? If we ignore for the
moment $H_{coul}$ it is obvious that the filled fermi sea \Vac\ is the
ground state of $H_{F,0}$ (for any given set of $n^a_F$ and $\l_a$, or
equivalently, for any given set of $\xi_a$'s, which is probably a more
appropriate specification of external background in our problem).  In
presence of $H_{coul}$ the argument is not so simple. Let us present
several independent reasons why the filled fermi sea should be the
ground state.  (a) In \cite{DMWQCDA}\ we have shown that the
expectation value of the meson bilocal operator $\M_{xy}(\l, t)$ (see
eq. (55) of \cite{DMWQCDA}) in this state provides the unique lowest
energy translation invariant solution to the classical equation of
motion. (b) The four-fermi interaction represented by $H_{coul}$ is
repulsive in nature. (c) It is easy to show that small number of
gauge-invariant mesonic excitations always {\em increase} the energy
of this state. (d) In the scaling that we describe in the next section
(involving $g$ and $L$) $H_{coul}$ is subleading to $H_{F,0}$ by a
factor of $1/N$. Thus, to leading order in $1/N$ the filled fermi sea
must be the ground state.

\subsection{Effective hamiltonian for gauge fields and
ground state of the full theory}

Now that we have computed the ground state of $H_F$ for fixed external
gauge field, let us construct the full ground state by the method of
separation of variables. What we will do now is similar in spirit to
solving the Schrodinger problem for a central force in three
dimensions where we look for solutions which are products of radial
and angular wave-functions. We solve the angular problem first and
find the eigenfunctions of the angular momentum operator. The
centrifugal term $L^2/r$ is evaluated by using the eigenvalue of the
angular momentum at a fixed $r$. This is then put back in the full
laplacian to derive an effective hamiltonian for the radial problem
which is then solved by appropriate methods.

The schematic correspondence between the above example and our problem
will be $ \psi \leftrightarrow (\theta, \phi)$, $\vec {\l}
\leftrightarrow r$, $H_F [\psi, \psi^\dagger, \vec \l] \leftrightarrow
L^2(\theta, \del_\theta, \del_\phi)/r$, $\Vac \leftrightarrow Y_{0,0}
(\theta, \phi)$ and  $H_{YM} \leftrightarrow - (1/r^2) \del_r (r^2
\del_r)$. Note the important fact that \Vac\ is independent of the
$\l_a$'s as is clear from the defintion \eq{fermisea}.

Let us write a product wavefunction of the system as
\be
| \Psi \rangle =  \Vac \otimes \Phi( \vec \l)
\label{fullstate}
\ee
Using \eq{hamiltonian}, \eq{efdef} and \eq{efvalue}\ we find that
\be
H | \Psi \rangle =   \Vac \otimes H_{\rm eff} \Phi ( \vec \l)
\ee
where
\be
H_{\rm eff} =  \sum_a {g\l_0 \over 4\pi } (- \del_{ \l_a}^2) +
E_F
\label{heff}
\ee
with $E_F$ given by \eq{efvalue}.

Our first guess at the full ground state would be $| \Psi \rangle =
\Vac \otimes \Phi_0 (\vec \l)$ where $\Phi_0(\vec \l)$ is the ground
state of $H_{\rm eff}$. However, as we have discussed in great detail
in \cite{DMWQCDA}\ such a state $| \Psi \rangle $ is not
gauge-invariant because of the shift \eq{fermilevelshift}\ of the
fermi levels under large gauge transformations. The correct
gauge-invariant ground state is given by a sum over fermi levels
\cite{DMWQCDA}
\be
| \Psi_0 \rangle = \sum_{ \vec n_F } \Vac \otimes
\Phi^{(0)}_{\vec n_F } ( \vec \l )
\label{fullgroundstate}
\ee
where the wavefunctions $\Phi^{(0)}_{ \vec n_F } ( \vec \l )$ are
given by
\be
\Phi^{ (0) }_{ \vec n_F } ( \vec\l ) =  u^{(0)} ( \vec \l + \vec  n_F )
\label{phiodef}
\ee
Here $u^{(0)} ( \vec \xi ) $, $\xi \in (-\infty, \infty)$ is the
ground state wave-function for the hamiltonian \eq{heff}\ where in
the kinetic term the operator $\del_{\l_a}$ has been replaced by
$\del_{\xi_a}$. In other words, $u^{(0)}(\vec \xi)$ satisfies the
differential equation
\be
{\cal H} u^{(0)} (\vec \xi) = E_0 u^{ (0) }( \vec \xi )
\label{ueq}
\ee
where
\be
{\cal H} = \sum_a {g\l_0 \over 4\pi} (-\del_{\xi_a}^2) + E_F (
\vec \xi)
\label{calhdef}
\ee
In the above $E_0$ denotes the lowest eigenvalue of the hamiltonian.
We shall discuss in the next section how to evaluate it.

Note that in expectation values computed in the full theory, the
implication of the sum-of-product structure \eq{fullgroundstate}\
of the full wave-function is that the effective range of the
eigenvalues becomes the real line.  Let us calculate, for instance,
the expectation value of the full hamiltonian $H$ in the state $|
\Psi_0 \rangle$.
\be
\langle \Psi_0 | H | \Psi_0 \rangle
= \sum_{ \vec n_F  } \prod_a \int_0^{\l_0} \! d\l_a \;
\Phi_{\vec n_F }^{ (0) *}( \vec \l )H_{\rm eff}
\Phi^{ (0) }_{\vec n_F }(\vec \l)
= \prod_a \int_{-\infty}^\infty \! d\xi_a \; u^{ (0) *}
( \vec \xi )  {\cal H} u^{ (0) }(\vec \xi)
\label{decompact}
\ee
where $H_{\rm eff}$ and ${\cal H}$ are given by \eq{heff}\ and
\eq{calhdef}. The last line shows that the presence of fermions forces
a {\em decompactification} of the eigenvalues $\l_a$ to the
gauge-invariant combination $\l_a + n^a_F
\l_0 = \xi_a$ which is a real number $\in (-\infty, \infty)$. It also
tells us that there is no fixed fermi level in a compact gauge theory
with dynamical gauge fields; rather, the effective theory of the gauge
fields $\vec \l $ in the fermi vacuum is given by a density matrix
$\rho(\vec n_F, \vec \l | \vec {n'_F}, \vec {\l'})$.  The sum over
$\vec {n_F}$ in the above equation corresponds to taking trace over
this density matrix.

\def\l{\mbox{$\bar \lambda$}}
\def\xib{\mbox{$ {\bar  \xi} $}}
\def\balph{\mbox{$ \bar \alpha $}}

\section{The Large-$N$ Expansion}
In this section we will discuss the large $N$ limit in detail and
present the $1/N$ expansion for some physical quantities. For
concreteness, we will consider the partition function
\be
\exp(- \beta F) \equiv Z = \Tr \exp( - \beta H)
\label{partition}
\ee
where $H = H_{YM} + H_F$ is given by \eq{hamiltonian}. We will also
evaluate towards the end of this section the vacuum expectation value
of the Wilson loop operator $\Tr U^m$ defined by \eq{wilsondef}.

The large $N$ limit involves defining a scaled coupling constant
\be
\bar g = g\sqrt N
\label{scaledgdef}
\ee
which is held fixed as $N \rightarrow \infty$. According to
\eq{lambdaperiod}, we must also define a scaled eigenvalue-period
\be
\lambda_0 = \sqrt N \l_0, \; \l_0 = { 2\pi \over \bar g L}
\label{scaledlambdao}
\ee
This necessitates scaling of $\l_a, \xi_a$ and $p_a$ to
\be
\l_a = \lambda_a /\sqrt N, \; \xib_a = \xi_a/ \sqrt N, \;
{\bar p}_a = p_a/\sqrt N
\ee
Note that in terms of the scaled variables ${\bar p}_a = -(i / N )
\del_{\l_a} $, implying that $\hbar$ is $1/N$. Let us, for example,
rewrite Eqns. \eq{ueq}\ and \eq{heff}\ for the ground state energy and
eigenfunctions in terms of the scaled variables
\be
{\cal H} u^{(0)} ( \{ \xib_a \} ) = E_0 u^{(0)} ( \{ \xib_a \})
\label{scaledueq}
\ee
\be
\ba
{\cal H} = {\cal H}_{YM} + {\cal H}_1 + {\cal H}_2 \\
{\cal H}_{YM} = (\bar g\l_0/ 4N \pi)\sum_a (-\del^2_{\xib_a}) \\
{\cal H}_1 = \bar g \lambda_0 \sum_a ( \xib_a/\lambda_0 + 1/2)^2 \\
{\cal H}_2 = (\bar g /4\pi N \l_0) \sum_{a,b} K_{\rm reg}([\xib_a
- \xib_b]/ \l_0)
\ea
\label{scaledheff}
\ee
where the $K_{\rm reg}$ has been defined in \eq{kernel} (note that in
\eq{kernel}\ $w_{ab}$ is also equal to $[\xib_a - \xib_b]/\l_0$).

\subsection{$\beta \to \infty$ limit and the ground state
energy}

In the $\beta \to \infty$ limit, the only state contributing to the
trace in \eq{partition}\ is the ground state, discussed in the
previous section. Thus the free energy $F$ simply coincides with the
ground state energy $E_0$ of \eq{scaledueq}.

Let us now discuss how to determine $E_0$. If the two-body interaction
${\cal H}_2$ was absent, $E_0$ would be simply given by
\be
E_{0,0} = {\bar g \over 2\sqrt{\pi}} N^{3/2}
\label{eoo}
\ee
{\em Proof:} The ground state for the $N$-fermion hamiltonian ${\cal
H}_{YM} + {\cal H}_1$ is given by the Slater determinant
\be
u^{0}_0 (\xib) = {1\over \sqrt{N !} } {\rm Det}_{i,j=0}^{N-1}
\phi_i(\xib_{j+1})
\label{uvalue}
\ee
where $\phi_n(\xib), n=0, \ldots, \infty $ are normalized
single-particle wavefunctions satisfying
\be
{\bar g \l_0 \over 2 \pi }
\big [- {\del^2_{\xib}  \over 2 N} + 2 \pi ({ \xib \over \l_0} + {1
\over 2})^2 \big] \phi_n(\xib) = \epsilon_n \phi_n(\xib)
\ee
It is easy to find explicit solutions of the above equation:
\be
\phi_{n}(\xib) = ({2^n n!\over \balph \sqrt{N \omega}})^{-1/2}
H_n(\sqrt{\omega} x) \exp(-\omega x^2/2), \;
\omega^2 \equiv 4 \pi
\label{phivalue}
\ee
where the variable $x$ is defined by
\be
(\xib/\l_0) + (1/2) = \balph x, \;
\balph \equiv  \l_0^{-1/2} N^{-1/4}
\label{xdef}
\ee
In the above $H_n$ are the standard Hermite polynomials. The energy
eigenvalues $\epsilon_n$ are given by
\be
\epsilon_n =  {\bar g \over \sqrt{\pi N} } (n + {1\over 2})
\quad n = 0, 1, \ldots
\ee
The ground state energy $E_{0,0}$ which is a sum over the first
$N$ energy levels clearly reproduces \eq{eoo}.

For later use let us also evaluate the density operator
\be
\rho(\xib) =  {1\over N} \sum_{i=1}^N \delta(\xib - \xib_a)
\label{densitydef}
\ee
in the state \eq{uvalue}. The result is \cite{MEHTA}
\be
\ba
\rho_{0,0}(\xib) = {1 \over N} \sum_{n=0}^{N-1}
|\phi_n(\xib)|^2 = 2 \balph \pi^{-3/4}
[(1 - x^2 { \sqrt{ \pi} \over N} )^{1/2}
\theta(1 - x^2 { \sqrt{ \pi} \over N} ) +
o(1/N)]
\ea
\label{densityvalue}
\ee

So far we have ignored the two-body interaction ${\cal H}_2$ in
\eq{scaledheff}\ (the additional subscript $0$ on the ground state
wavefunction and energy denotes this fact). We will see shortly that
in the only sensible scaling available in the theory this term will be
of lower order in $N$. In other words, the correction terms to
\eq{eoo}\ coming from the two-body interaction will turn out to be
$1/N$ lower order than $N^{3/2}$, leading to the result that the total
ground state energy is $E_0 = N^{3/2} \bar g / (2 \sqrt{\pi}) +
o(N^{1/2})$.

\vspace{2 ex}

\noindent\underline{Collective field theory}

\vspace{2 ex}

Since the result $E_0 \propto N^{3/2}$ is rather unexpected (naively
one would expect the pure Yang-Mills result $N^2$), let us try to
understand this from some simple scaling analysis. The easiest
framework to do such an analysis is collective field theory
\cite{JEVSAK}.  The hamiltonian ${\cal H}$ in \eq{scaledheff}\
corresponds to the following collective field theory hamiltonian (the
subscript `$c$' stands for collective)
\be
\ba
\Hc =  \Hc_{YM} +  \Hc_1 +  \Hc_2
\\
\Hc_{YM} = N^2 (\bar g \l_0 / 2\pi) \;
\int d\xib \, (\rho(\xib) / 2) [\Pi^2(\xib) +
{ \pi^2 \over  3} \rho^2(\xib) ]
\\
\Hc_1 = N (\bar g \l_0 / 2 \pi) \;  \int  d\xib \,
\rho(\xib) [ 2 \pi (\xib/\l_0 + 1/2)^2 ]
\\
\Hc_2 = N (\bar g \l_0/ 2 \pi) \; (1 / 2\l_0^2)
\int d\xib \,  \int d\xib' \,  K_{\rm reg}(
\xib - \xib'/\l_0) \rho(\xib) \rho(\xib')
\ea
\label{hcdef}
\ee
Here $\rho(\xib)$ is the eigenvalue density defined in \eq{densitydef}.
$\Pi(\xib)$ is defined such that $\Pi(\xib) \rho(\xib)$ corresponds to
the momentum density: $ \int d\xib\, \Pi(\xib) \rho(\xib) = (1/N) \sum_a
{\bar p}_a$. This implies the following commutation relation
\be
[ \rho(\xib), \Pi(\xib') ] = - { i \over N^2} \del_{\xib} \delta(\xib -
\xib')
\label{commutator}
\ee
The theory is defined with the constraint
\be
\int d\xib \rho(\xib) = 1
\label{rhoconstraint}
\ee
which is a consequence of the definition \eq{densitydef}.

\vspace{2 ex}

\noindent\underline{Classical analysis}

\vspace{2 ex}

Let us regard the collective field hamiltonian $\Hc$ as a function of
the classical variables $\rho(\xib), \Pi(\xib)$. The Poisson bracket
between them is simply \eq{commutator}\ without the $i$ on the right
hand side
\be
[ \rho(\xib), \Pi(\xib') ]_{PB} = - {1 \over N^2} \del_{\xib}
\delta(\xib - \xib')
\label{poisson}
\ee
This leads to the following equations of motion
\be
\ba
\del_t \rho(\xib) = - {\bar g \l_0 \over 2 \pi}
\del_{\xib} (\rho(\xib) \Pi(\xib)) \\
\del_t \Pi(\xib) = - {\bar g \l_0 \over 2 \pi }
   \del_{\xib} [{\Pi^2 \over 2} +  {\pi^2 \over 2}\rho^2 +
{ 2 \pi \over N} (\xib/\l_0 + 1/2)^2 + \\
\,~~~~~~~~~~~~~~~~~~~~~~~~~~~~~~~~~~~~~~~{ 1 \over
N \l_0^2 } \int d\xib' \, \rho(\xib') K_{\rm reg}((\xib
- \xib')/\l_0) ]
\ea
\label{eqm}
\ee
For time-independent solutions these reduce to
\be
\del_{\xib} [ {\pi^2 \over 2} \rho^2 +
{ 2 \pi \over N} (\xib/\l_0 + 1/2)^2 + { 1 \over
N \l_0^2 } \int d\xib'\, \rho(\xib') K_{\rm reg}((\xib
- \xib')/\l_0) ] = 0
\label{staticeqm}
\ee
If we take the naive $N \to \infty$ limit, the potential and
interaction terms drop out and we get a constant density
\be
\rho(\xib) = \hbox{constant}
\label{naivestatic}
\ee
However, since that the range of $\xib$ is noncompact (due to the
phenomenon of decompactification discussed in the last section) such a
constant density is unnormalizable, that is, it cannot satisfy
\eq{rhoconstraint}. There are no gauge-invariant cutoffs available on
the range of $\xib$ either which can save the situation.

Clearly in order to obtain normalizable solutions for $\rho(\xib)$
the large $N$ limit must be taken in such a way that one or both of
the terms $\Hc_1$ and $\Hc_2$ in \eq{hcdef}\ are of the same
order as $\Hc_{YM}$.  Now it is easy to see that only $\Hc_2$ by
itself, being translation invariant, cannot produce
localization. Indeed, if one includes in \eq{staticeqm}\ terms coming
from $\Hc_{YM}$ and $\Hc_2$ and drops those coming from $\Hc_1$, the
solution is still $\rho(\xib) =$ constant, which is untenable. Thus
we must ensure that the simple harmonic potential term in
\eq{staticeqm}\ survives in the large $N$ limit (irrespective of what
happens to the interaction term).  In other words, in \eq{hcdef}\
$\Hc_{YM}$ and $\Hc_1$ must be of the same order.

The clue to how the above can be achieved is provided by the formula
\eq{densityvalue}. This suggests that the correct coordinate in terms
of which a sensible $N$ scaling of the collective field theory may be
available is not $\xib$, but $y$, defined by
\be
x = \sqrt{N} y, \quad \hbox{equivalently} \quad
{\xib \over \l_0} + {1 \over 2} = \alpha y, \qquad
\label{ydef}
\ee
where
\be
\alpha = {\l_0}^{-1/2} N^{1/4} \equiv \balph \sqrt N
\label{alphadef}
\ee
The new density variable $\rhot(y)$ is given by
\be
\rhot(y) = \rho(\xib) d\xib/dy = \rho(\xib)  N^{1/2} \alpha^{-1}
\label{rhotdef}
\ee
Note that the density expectation value \eq{densityvalue}\ in terms
of the new variable reduces to
\be
\rhot_{0,0} (y) \equiv \rho_{0,0} (\xib) N^{1/2} \alpha^{-1}
= 2 \pi^{-3/4} ( 1 - y^2 \sqrt \pi)^{1/2} \theta( 1 - y^2 \sqrt \pi )
\label{rhotvalue}
\ee
which is an $N$-free expression. We also define a new momentum
variable $\Pit(y)$ by
\be
\Pit(y) = \Pi(\xib) N^{1/2} \alpha^{-1}
\label{pitdef}
\ee
such that the Poisson bracket is kept invariant. This also ensures
that $N$ scales out of the combination $\Pi^2(\xib) + \rho^2 ( \xib )
/ 12 $ in $\Hc_{YM}$, Eqn. \eq{hcdef}. In terms of these, the collective
field hamiltonian becomes
\be
\ba
\Hc_{YM} = N^{3/2}\,(\bar g / 2 \pi) \;  \int dy   \,
(\rhot(y)/2) [\Pit^2(y) +  (\pi^2/3) \rhot^2(y) ]
\\
\Hc_1 = N^{3/2}\, (\bar g / 2 \pi) \; \int dy \, \rhot(y)\, 2 \pi y^2
\\
\Hc_2 = N^{1/2}\, (\bar g / 2 \pi) \alpha^2 \; \int dy \, \int dy' \,
\rhot(y) \rhot(y') K_{\rm reg}(\alpha(y -y'))
\ea
\label{scaledhc}
\ee

There are several remarks to be made here. First of all, note that
$\Hc_{YM}$ and $\Hc_1$ scale as $N^{3/2}$. It is easy to show that if
we ignore the $\Hc_2$ piece, then the time-independent equations
(recall \eq{staticeqm}) in the new variables have a solution
$\rhot(y)$ which exactly coincides with \eq{rhotvalue}\ and the the
classical energy of this configuration is identical to
\eq{eoo}. The second point is, $\Hc_2$ is subleading compared to
the first two terms in $\Hc$ by $1/N$, if we demand that $\alpha$ is
$N$-independent as $N \to \infty$. One can therefore treat this term
as a perturbation. The reason for demanding $N$-independence of
$\alpha$ is that that is the only way a scaled form of $\Hc_2$ can be
obtained. Indeed one can easily rule out $N$-dependence of $\alpha$ on
other physical grounds also. If, for example, $\alpha$ grew with $N$,
the function $K_{\rm reg}(\alpha(y - y'))$ would become infinitely
discontinous (its poles become infinitely dense). If, on the other
hand, $\alpha$ went as some inverse power $N^{-\gamma}, \gamma > 0$,
then by using the formula
\be
K_{\rm reg}(\alpha y) = \sum_{m=1}^\infty \zeta(2m + 1) (\alpha
y)^{2m}
\label{kexpansion}
\ee
one can see that although from the point of view of convergence the
$\Hc_2$ perturbation series would be sensible, it would have the
unphysical feature that different parts of the two-body interaction
contribute in different orders of perturbation theory leading to an
incorrect representation of the nature of the interaction.

\newpage

\noindent\underbar{Scaling of length}

\vspace{2 ex}

Eqn. \eq{alphadef}\ is equivalent to
\be
L = {\alpha^2 2 \pi \over \bar g} N^{-1/2}
\label{lengthscaling}
\ee
In view of the fact that $\alpha$ is $N$-independent, we see that as
$N \to \infty$, the (bare) length $L$ of the space circle goes to zero
as $L = N^{-1/2} \bar L$ where $\bar L$ is held constant. The
parameter $\alpha$ can be identified as $ \sqrt{\bar g \bar L/ 2
\pi}$.

\vspace{2 ex}

\noindent\underline{Scaling of time and energy}

\vspace{2 ex}

Let us rewrite the classical equations of motion using the variables
$\rhot(y)$, $\Pit(y)$. We get
\be
\ba
N^{1/2} \, \del_t \rhot(y) = - (\bar g / 2 \pi)
\del_y (\rhot(y) \Pit(y))
\\
N^{1/2} \, \del_t \Pit(y) = - (\bar g / 2 \pi)
\del_y [ {\Pit^2 \over 2} + {\pi^2 \over 2} \rhot^2  + 2 \pi y^2 +
{\alpha^2 \over N} \int dy' \,
\rhot(y') K_{\rm reg}(\alpha(y - y'))]
\ea
\label{scaledeqm}
\ee
It is clear that $N$ scales out of the classical equations if we work
in terms of a scaled time $\bar t$ defined by
\be
t = N^{1/2} \bar t
\label{tscaling}
\ee
Now a scaling of time (in the Euclidean framework a scaling of
$\beta$) implies an inverse scaling of energy
\be
E = N^{-1/2} \bar{E}
\label{escaling}
\ee
In other words, $\exp ( i H t) = \exp ( i \bar{H} \bar{t}),
\exp (- \beta H ) = \exp ( -\bar{\beta} \bar{H}).$

In this new scaling, the collective field hamitonian becomes
\be
\bar {\Hc} = {N^2 \bar g  \over 2 \pi}
\int dy \, {\rhot(y) \over 2} \big[ \Pit^2 + {\pi^2 \over
3} \rhot^2 + 4 \pi y^2  + {\alpha^2 \over  N} \int dy' \,
\rhot(y') K_{\rm reg} (\alpha(y - y')) \big]
\label{newscaledhc}
\ee
The scaled ground state energy is, therefore,
\be
\bar {E_0} = N^2 {\bar g \over 2 \sqrt \pi} [ 1 + o(1/N)]
\ee
In the following we will study how the $1/N$ corrections follow from
a WKB analysis.

\vspace{2 ex}

\noindent\underline{$1/N$ corrections and WKB}

\vspace{2 ex}

The commutation relation between $\rhot(y)$ and $\Pit(y)$ is given by
(recall \eq{commutator})
\be
[ \rhot(y), \Pit(y') ] =  - {i \over N^2} \del_y \delta(y - y')
\label{scaledcommutator}
\ee
which implies that $\Pit(y)$ can be represented by the following
differential operator representation
\be
\Pit(y) = - {i \over N^2} \del_y {\delta \over \delta \rhot(y)}
\label{momdiffop}
\ee
To construct a WKB solution we consider wave-functions of the form
\be
\Psi [ \rhot ] = \exp \big( i N^p S [ \rhot ] \big)
\label{wkb}
\ee
where we have kept the real number $p$ unspecified and will determine
it by demanding a scaled time-independent Schrodinger equation. The
latter is given by
\be
\bar {\Hc} \Psi [ \rhot ] = \bar E \Psi [ \rhot]
\label{wkbenergydef}
\ee
where $\bar {\Hc}$ is given by \eq{newscaledhc}.
Using the above differential
operator representation of $\Pit(y)$, \eq{momdiffop}, we get the
following equation determining $S [ \rhot ]$ in terms of $
\bar E$:
\be
\ba
\bar E = {N^2 \bar g \over 2 \pi} \int dy { \rhot \over 2}
\big[ N^{2p -4} (\del_y {\delta S \over \delta \rhot(y)})^2
- i N^{p - 4} (\del_y {\delta \over \delta \rhot(y)})^2 S
+ {\pi^2 \over 3}\rhot^2 + 4\pi y^2 + \\
\,~~~~~~~~~~~~~~~~~~~~~~~~~~~~ {\alpha^2  \over N} \int dy'
\rhot(y') K_{\rm reg}(\alpha(y - y'))  \big]
\ea
\label{riccati}
\ee
In order to get a scaled Hamilton-Jacobi equation in the leading $N$
order we must put $p=2$. This implies the following perturbation
series for $S$
\be
\Psi[\rhot] = \exp [ i N^2 S], \qquad S = S_0 + {1\over N} S_1
+ {1 \over N^2} S_2 + \ldots
\ee
It is easy to write down from \eq{riccati}\ equations determining
$S_n$ in terms of the lower order terms.

The above analysis also implies that the energy $\bar {E_0}$ for the
ground state should receive corrections to \eq{eoo}\ as follows
\be
\bar {E_0} = N^2 \bigg[ {\bar g  \over 2 \sqrt\pi} + {1\over N}
\Delta_{0,1} + {1\over N^2} \Delta_{0,2}  + \ldots \bigg]
\label{eseries}
\ee
We have explicitly verified this by computing the many-body
perturbation theory diagrams of the hamiltonian \eq{hcdef}.

\vspace{2 ex}

\noindent\underline{Introduction of multiple flavours}

\vspace{2 ex}

In the above we have considered the case of a single flavour, $n_F =
1$.  It is easy to extend the above calculations to multiple flavours.
The essential new physics point is the following. The dirac sea $| F
\rangle $ for $n_F > 1$ flavours of free fermions is given by
\be
\ba
\psi^a_{i,n} | F \rangle = 0 \; \hbox{for} \; n > n^{a,i}_F\\
\psi^{a\dagger}_{i,n} | F \rangle = 0 \; \hbox{for} \; n \le n^{a,i}_F
\ea
\label{nfsea}
\ee
where $\psi$'s above represent the right-handed fermions. The
left-handed fermions represent similar equations corresponding to
the filling of all the levels starting from $n^{a,i}_F + 1$ {\em
upwards}.

If the fermi levels for different flavours are different, then one
would anticipate several conceptual difficulties. For instance, if
there is no notion of a single $n^a_F$ how does one construct $\xib_a =
\lambda + n^a_F \l_0$ to describe the decompactified effective theory
of the eigenvalues? Fortunately, because of the chiral $U(n_f) \times
U(n_f)$ symmetry of the hamiltonian, we must demand that the ground
state is a flavour singlet. This happens only if the $n^{a,i}_F$s
above are all independent of $i$ (this can be proved by operating
the flavour rotation generators on $| F \rangle$). Having said
that, we can again use the notation \Vac\ instead of $| F \rangle$
to represent the dirac sea.

Using the above, it is easy to show that the dirac sea is again an
eigenstate of the hamiltonian upto $1/N$ corrections. In other words
\eq{efdef}\ is again valid, except that the energy eigenvalue $E_F$
is now $n_f$ times the expression \eq{efvalue}. As a result in the
calculation of the ground state energy according to \eq{scaledheff},
$\Hc_{YM}$ remains the same but $\Hc_1$ and $\Hc_2$ get multiplied by
$n_f$. It is simple to show that the ground state energy is now given
by
\be
\bar {E_0} = N^2 \big[ \sqrt{n_f} {\bar g  \over 2 \sqrt \pi}
+ o(1/N) \big]
\label{enf}
\ee
The eigenvalue density is given by
\be
\rhot(y) = 2 n_f^{1/4} \pi^{-3/4} ( 1 -  y^2 \sqrt{\pi n_f})^{1/2}
\theta( 1 - y^2 \sqrt{\pi n_f}) + o({ 1 \over N})
\label{rhonf}
\ee
Eqn. \eq{enf}\ clearly shows that adding fermions affects the
leading $N$ behaviour of the vacuum energy.

\subsection{Finite $\beta$ partition function and Excitations}

The method presented above can be generalized to $\beta \ne 0$.  In
this case, one needs to consider excited states in the fermionic
theory. This leads to subleading corrections (in $1/N$) to the
effective hamiltonian for the gauge fields. The free energy can again
be computed. We will not present here the explicit result but merely
state that the $N$-scalings presented above also work at finite values
of $\beta$ and that the leading term (in $N$) in the expression for
the free energy still has a non-trivial dependence on $n_f$.

\subsection{Vacuum expectation value of the Wilson loop operator}

In order to understand how the introduction of dynamical fermions
affects the leading large $N$ behaviour of various quantities, it is
instructive to calculate vacuum expectation value $ \langle W_m
\rangle = \langle \Tr U^m \rangle $ of the Wilson loop operator
(recall \eq{wilsondef}\ and \eq{gaugefixedwilson}). Using the
expression for the ground state wavefunction \eq{fullgroundstate}\ and
the definition \eq{densitydef}\ of the density operator it is easy to
show that
\be
\langle W_m \rangle =
\int d\xib \langle \rho(\xib) \rangle e^{ i 2 \pi m \xib / \l_0}
\label{wilsonvaluea}
\ee
On the right hand side the expectation value is meant in the ground
state of the effective theory of eigenvalues defined by
\eq{scaledueq}\ and \eq{scaledheff}, and is given by \eq{rhonf}.
Using \eq{ydef}\ to express the exponent in terms of $y$ we get
\be
\langle W_m \rangle =  2 (-1)^m  (1 + { \del^2 \over \del {x_m}^2 } )
J_0(x_m) + o(1/N), \quad x_m \equiv 2 \pi^{3/4} \alpha { m \over
n_f^{1/4}}
\label{wilsonvalue}
\ee
where $J_0$ denotes the Bessel function of order zero.

Let us compare this with the result in pure Yang-Mills theory on a
two-dimensional cylinder. The latter theory is described by the path
integral \eq{gaugefixedz}\ with no fermion integration and $S_F
=0$. It is easy to show in this case the eigenvalue density is a
constant (which, unlike with fermions, is perfectly allowed because
the range of eigenvalues remains compact) and therefore
\be
\langle W_m \rangle_{YM} = \delta_{n,0} + o( { 1 \over N} )
\label{wilsonvalueym}
\ee
One can also recover this by taking $n_f \to 0$ in \eq{wilsonvalue}.

It is clear that the introduction of fermions changes the result
for $\langle \Tr U^m \rangle $ in the leading $N$ order.

\section{Concluding Remarks}
To conclude we remark that QCD$_2$ on the cylinder shows considerable
amount of subtlety and surprise. The main surprise is that the leading
term in a $1/N$ expansion for the vacuum energy and for the {\em vev}
of the Wilson loop operator is changed by the existence of fermions,
contrary to what one would expect by standard $N$-counting
arguments. The essential reason for this is that coupling to quarks
leads to a `decompactification' of the eigenvalues. This effect is not
perturbative and persists even at $N= \infty$, thus changing the
leading behaviour of pure Yang-Mills theory. Indeed it is a very
interesting question what the present analysis can tell us about $3+1$
dimensional gauge theories in a compact space. In particular it would
be interesting to know whether the phenonmenon of decompactification
persists in four dimensions and whether as a result the pure
Yang-Mills theory is affected by inclusion of quarks in the leading
$N$ order.

\vspace{3 ex}

\noindent{\bf Acknowledgements:}  We would like to thank David Gross
for critical comments and useful discussions.

\end{document}